# Chapter 16: Photonic molecules and spectral engineering

Svetlana V. Boriskina

Department of Electrical and Computer Engineering, Boston University, Boston MA

**Abstract**

This chapter reviews the fundamental optical properties and applications of photonic molecules (PMs) - photonic structures formed by electromagnetic coupling of two or more optical microcavities (photonic atoms). Controllable interaction between light and matter in photonic atoms can be further modified and enhanced by the manipulation of their mutual coupling. Mechanical and optical tunability of PMs not only adds new functionalities to microcavity-based optical components but also paves the way for their use as testbeds for the exploration of novel physical regimes in atomic physics and quantum optics. Theoretical studies carried on for over a decade yielded novel PM designs that make possible lowering thresholds of semiconductor microlasers, producing directional light emission, achieving optically-induced transparency, and enhancing sensitivity of microcavity-based bio-, stress- and rotation-sensors. Recent advances in material science and nanofabrication techniques make possible the realization of optimally-tuned PMs for cavity quantum electrodynamic experiments, classical and quantum information processing, and sensing.

## 16.1 Introduction

As other chapters in this volume clearly demonstrate, optical micro- and nanocavities offer tremendous opportunities in creating, studying and harnessing confined photon states [1-6]. The properties of these states are very similar to those of confined electron states in atoms. Owing to this similarity, optical microcavities can be termed 'photonic atoms'. Taking this analogy even further, a cluster of several mutually-coupled photonic atoms forms a photonic molecule [7]. As shown in Fig. 16.1, typical PM structures consist of two or more light-confining resonant cavities such as Fabry-Pérot resonators, microspheres, microrings, point-defect cavities in photonic crystal (PC), etc [8]. The first demonstration of a lithographically-fabricated photonic molecule (Fig. 16.1a) was inspired by an analogy with a simple diatomic molecule [7]. However, other nature-inspired PM struc-tures (such as 'photonic benzene') have been proposed and shown to support confined optical modes closely analogous to the ground-state molecular orbitals of their chemical counterparts [9]. The most celebrated example of a coupled-cavity structure is a coupled-resonator optical waveguide (CROW), which is formed by placing photonic atoms in a linear chain [10]. The energy transfer along the chain can be achieved through nearest-neighbor interactions between adjacent cavities (photon hopping), and the unique dispersion characteristics of CROWs make possible the realization of ultra-compact on-chip optical delay lines [11-13]. Optical properties of more complex PM structures considered in this chapter depend on mutual coupling between all the cavities forming the PM, and can be optimally-tuned by adjusting the sizes and shapes of individual cavities as well as their positions. In this chapter, we first review the mechanism of mode coupling and splitting in microsphere, microdisk, and point-defect diatomic PMs, and introduce classification of the PM super-modes. We then demonstrate various ways of engineering the PM super-modes spectra, and explore the opportunities of post-fabrication tuning and dynamical modulation of PM structures.

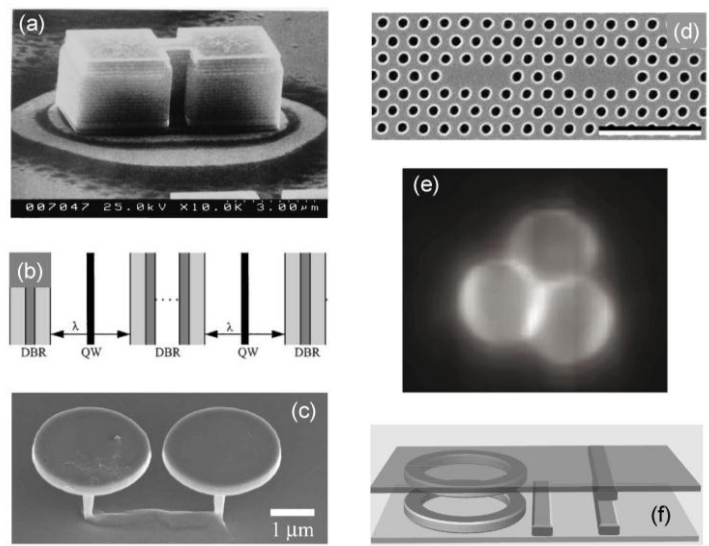

**Fig. 16.1** Typical configurations of PMs: (a) square-shape photonic dots coupled via a semiconductor bridge [7], (b) planar Fabry-Pérot cavities coupled through a partially-transparent Bragg mirror [14], (c) whispering-gallery mode microdisks side-coupled via an airgap [15], (d) closely located defects in a photonic crystal membrane [16], (e) triangular photonic molecule composed of touching microspheres [17], (f) vertically coupled microrings [18].

Unique optical properties of photonic atoms, including light confinement in compact structures that enable modification of the optical density of states and enhancement of nonlinear material properties, ultra-high quality factors and sensitivity to environmental changes have made them attractive building blocks for a variety of applications in basic science, information processing, and biochemical sens-ing. Mechanical tunability and increased design flexibility offered by PM structures make possible not only enhancing but also adding new functionalities to microcavity-based optical devices. For example, a proposal to use the effect of position-dependent splitting of whispering gallery (WG) modes in evanescently-coupled microsphere resonators as a basis of a high-sensitivity coordinate meter dates back to 1994 [19], even before the term 'photonic molecule' was introduced. In recent years, PM structures have also entered the field of sensing as optical transducers for high-sensitivity stress [20], rotation [21-23], and refractive index [24, 25] measurements. Lineshape- and bandwidth-tuning capabilities of PM structures drive their applications as optical filters and switches [26-30] and also help to increase sensitivity of PM-based sensors [31, 32]. Furthermore, the optical interactions between photonic atoms may be tuned to enhance select modes in PM structures and to shape their angular emission profiles [8, 33-36], paving the way to realizing low-threshold single-mode microlasers with high collection efficiency. It should be noted that many of the above-mentioned applications of PMs to be discussed in this chapter have well-known analogs in the field of microwave and millimeter-wave engineering [37-41]. However, it was recently discovered that PM structures may also serve as simulators of quantum many-body physics, yielding unique insights into new physical regimes in quantum optics and promising applications in quantum information processing [42]. A few examples of these exciting new opportunities will also be discussed [14, 43-47].

## 16.2 Optical properties

If individual photonic atoms are brought into close proximity, their optical modes interact and give rise to a spectrum of PM super-modes. Adopting the terminology used in the studies of localized plasmonic states coupling, this mode transition and splitting can be called mode hybridization [48].

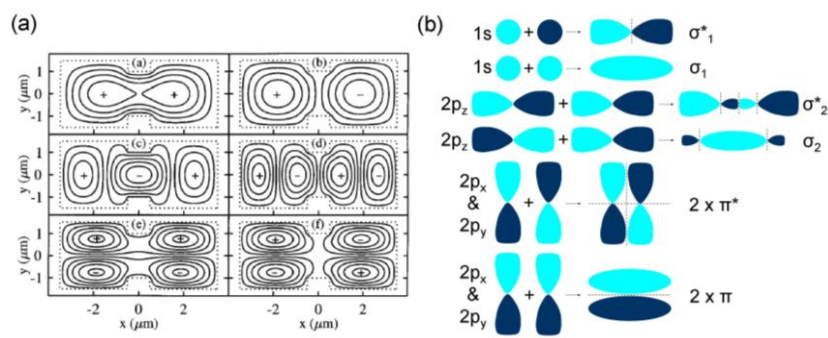

**Fig. 16.2** (a) Calculated electric field distributions of the six lowest-energy super-modes in the two-atom PM shown in Fig. 1(a) [7]; (b) Diagram of bonding and anti-bonding molecular orbitals in a diatomic chemical molecule.

The electric field patterns of the six lowest-energy super-modes in the double-cavity PM shown in Fig. 16.1(a) are plotted in Fig. 16.2(a). The super-modes can be characterized by their parities, either even (++) or odd (+−) with respect to the PM axes of symmetry. The observed splitting of localized modes of individual photonic atoms into PM super-modes is analogous to that of the electron states of a diatomic molecule. For example, the even-(odd-)parity super-modes shown in the top row in Fig. 16.2(a), which arise from the constructive (destructive) interference of lowest-energy non-degenerate modes in individual photonic atoms, correspond to bonding and anti-bonding σ-like molecular orbitals (MO) formed by superposition of *s* atomic orbitals [see Fig. 16.2(b)]. The other four electric field patterns in Fig. 16.2(a), which correspond to σ-like and π-like MOs formed by superposition of *p* atomic orbitals, are the result of the interference of two double-degenerate higher-energy optical modes of individual photonic atoms.

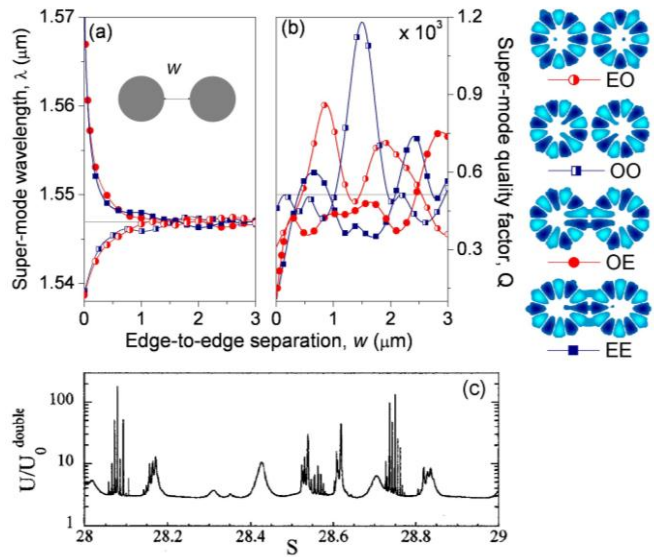

**Fig. 16.3** Splitting of WG$_{6,1}$ modes in a double-microdisk PM: wavelengths (a) and quality factors (b) as a function of the airgap width between microdisks (TE polarization, disks radii 0.9 μm, effective refractive index *n*=2.63). The inset shows the magnetic field distributions of bonded and anti-bonded WG super-modes. (c) Bi-sphere PM internal energy spectrum normalized to the vacuum electromagnetic energy within the sphere ($U_0^{double}=(2/3)r^3$) vs spheres size parameter (*S=kr*). The PM is composed of two identical touching spheres with *n*=1.59 excited by the plane wave incident at 10° to the PM major axis [49].

Electromagnetic coupling between point-defect PC cavities [16, 20] as well as WG mode microdisk, microring, and microsphere resonators [19, 49-57] also result in the splitting of individual cavity modes into blue-shifted anti-bonding super-modes and red-shifted bonding super-modes (see the super-mode field distributions on the right of Fig. 16.3). The shift of the resonant frequency depends on the strength of the coupling between individual atoms. Furthermore, the values of the super-mode Q-factors also change (which may be referred to as 'loss splitting' [16]) and vary with the change of the inter-cavity coupling efficiency. These two effects are illustrated in Fig. 16.3(a,b) for the case of two microdisk resonators laterally coupled via an airgap. WG modes in individual microdisks ($WG_{mn}$) are double degenerate and are classified by two indices that correspond to the number of azimuthal ($m$) and radial ($n$) field variations, respectively. Breaking the symmetry of the structure results in lifting of the WG mode degeneracy, however, bonding and anti-bonding super-modes form closely-located doublets in the PM optical spectrum, and may be indistinguishable experimentally [53]. It should be noted that PMs of symmetrical configurations can support both non-degenerate and degenerate super-modes [8, 18, 25, 33, 35]. Bonding modes, which have enhanced field intensity in the inter-cavity gap, are more sensitive to the gap width than anti-bonding ones, which have a nodal plane (a plane of zero intensity) between the microcavities (see Fig. 16.3(a)). Fig. 16.3(a) also demonstrates that, differently from splitting of electronic energy levels in chemical molecules, the parity of the blue- and red-shifted PM super-modes is not conserved with the change of the coupling distance. Oscillations of the super-mode resonant frequencies around that of the isolated cavity, reflect the oscillatory behavior of the modes evanescent field.

In microsphere resonators, the WG-modes are classified by angular ($l$), azimuthal ($m$, $-l \leq m \leq l$), and radial ($n$) mode numbers, and are ($2l+1$)-fold degenerate. The most interesting for practical applications are the fundamental modes (with $n = 1$ and $m = l$), which feature the highest Q-factors and the smallest mode volumes. Optical coupling between microspheres lifts the WG mode degeneracy and leads to a multi-peak spectral response of the PM structures [49, 52, 54, 55, 58, 59]. The spectrum of the internal energy in a bisphere PM, calculated by using a tight-binding approximation [49], is presented in Fig. 16.3(c) and shows a fine structure of the WG super-modes. The fine structure of the bisphere super-modes has also been observed experimentally by studying the photoluminescence from melamine-formaldehyde spherical microcavities with a thin shell of CdTe nanocrystals [52, 58]. In some situations, only two broad peaks can be observed in a bisphere measured spectrum (as in the case of coupled microdisk resonators), which represent broadened envelopes of unresolved multiple resonances [60].

Intermixing between WG super-modes in microdisk and microsphere PM structures that are characterized by different values of radial and azimuthal quantum numbers can also occur [61, 62]. For example, if the wavelengths of the first- and second-radial-order WG super-modes have a crossing point when plotted as a function of a tuning parameter (e.g., airgap width), the induced anti-crossing coupling between these modes produces PM super-modes with greatly reduced Q-factors [61]. Furthermore, even in the case of 'weak' mode coupling (when the modes dispersion curves do not cross each other) the interband coupling can significantly modify the super-mode dispersion curves [62].

If $N$ photonic atoms are side-coupled to form a linear chain, the optical spectrum of the resulting PM features multiple super-mode peaks, with the number of peaks proportional to the number of photonic atoms and to the degeneracy number of the modes in the stand-along atom. This situation is illustrated by Fig. 16.4(a), which shows the splitting of WG modes in a finite-size linear chain composed of 7 identical microdisks with effective refractive indices $n_{eff} = 2.9$ and diameters $d = 2.8$ μm with the change of the width of the airgaps between the disks. In the structures composed of a few strongly-coupled resonators such as that shown in Fig. 16.4(a), the modes splitting can be resolvable experimentally [53]; however, excessive mode splitting in long microcavity chains results in the appearance of discrete photonic bands [10, 63-67]. This effect (which is demonstrated in Fig. 16.4(b) [63]) is analogous to the appearance of electronic bands in semiconductors as a result of the overlap of atomic wave functions and multiple atomic level splitting. In the limit of an infinite number of coupled photonic atoms (infinite-length CROWs), the discrete band becomes a continuum, making possible an analytical description of the CROW dispersion properties [10]. It has also been shown, both theoretically and experimentally, that coupling of many photonic atoms in chains [10, 68, 69] or two-dimensional (2D) arrays [70, 71] can yield flat optical bands, which enable slow light engineering and enhancement of light-matter interactions.

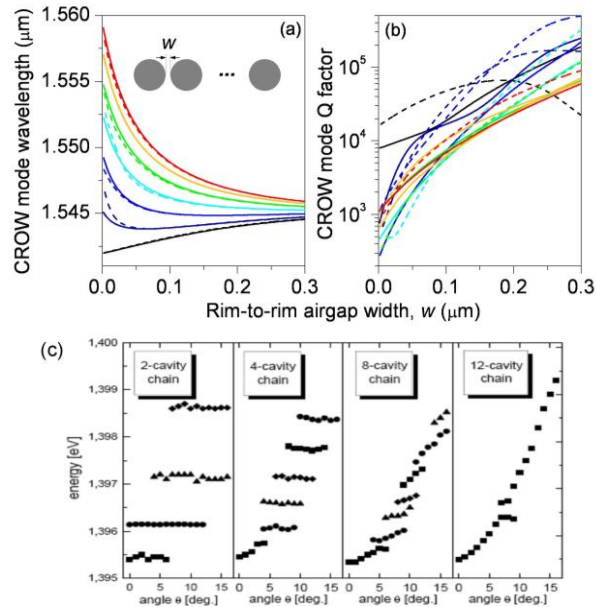

**Fig. 16.4** Splitting of WG$_{12,1}$ modes in a linear chain of seven 2.8μm-diameter microdisks with the decrease of the inter-cavity airgap width w: (a) shift of resonant wavelengths and (b) degradation of the mode Q-factors [72]; (c) Experimentally observed optical band formation in linear chains of coupled square-shaped photonic atoms: energies of optical modes versus the angle along the chains for varying numbers of coupled cavities [63].

If an isolated photonic atom is driven by external excitation field (plane wave, dipole or optical waveguide evanescent field), its spectral response usually features a series of Lorentzian peaks corresponding to the excitation of its morphology-dependent resonances. Coherent interference of optical modes in PMs makes possible the manipulation of the bandwidth and/or shape of the resonant peaks. One interesting phenomenon that can be observed in PM structures composed of two or any even number of photonic atoms is coupled-resonator-induced transparency (CRIT), or resonant cancellation of absorption caused by mode splitting and destructive interference [60, 73-75]. Once again, an analogy can be drawn between PMs and atomic systems, where a phenomenon of electromagnetically-induced transparency (EIT) can be observed under coherent excitation by an external laser. EIT in atomic systems is a result of destructive quantum interference of the spontaneous emission from two close energy states that reduces light absorption (see Fig. 16.5(a)). CRIT is manifested as a sharp dip in the PM absorption spectrum as shown in Fig. 16.5(b) (or, alternatively, a sharp peak in the PM transmission spectrum) caused by the destructive interference between PM optical modes. Unlike EIT effect, which can only be achieved by making use of limited number of accessible transition frequencies and requires a high intensity drive laser, CRIT is a result of classical optical interference. Thus, a need for a powerful drive laser is eliminated and the PM spectral lineshape can be flexibly manipulated by tuning the PM geometry (e.g., the inter-resonator coupling efficiency can play the role of the strength of the pump field in EIT).

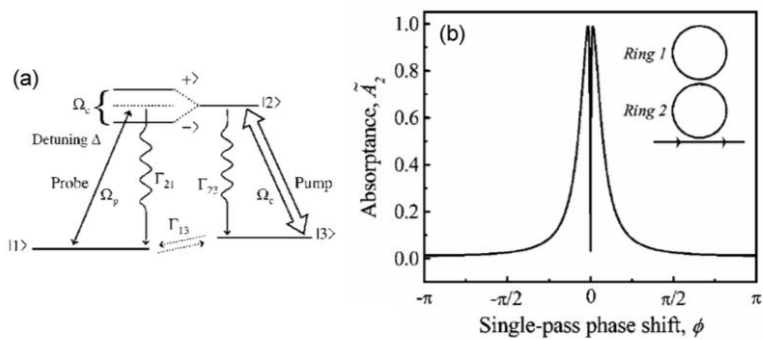

**Fig. 16.5** (a) EIT in a three-level atomic system: destructive interference of the transition probability amplitudes between two nearly-degenerate energy states (split by a strong resonant pump) and a common ground state (see [73] for detail), (b) CRIT in a diatomic PM structure: absorptance in the PM structure excited by a bus waveguide exhibits a sharp dip when plotted as a function of the single-pass phase shift (an analog of the frequency detuning) [73].

It should also be noted that CRIT can be realized not only in PM structures composed of identical photonic atoms as shown in Fig. 16.5(b), but also in asymmetrical PMs composed of size-mismatched optical cavities [73, 76]. In this case, the cavity sizes and the inter-cavity gap width should be adjusted to ensure efficient coupling between their optical modes. The strength of the optical coupling between modes of individual microcavities depends on the difference between their resonant wavelengths. For example, the coupling strength between two side-coupled microdisk resonators can be tuned by changing the diameter of one disk while keeping the diameter of the other fixed, as shown in Fig. 16.6 [77]. It can be seen that when the wavelengths are tuned closer to each other, the PM supermodes strongly couple with the avoided crossing scenario [Fig. 16.6(a)]. Frequency anti-crossing is accompanied by crossing of the corresponding widths of the resonance states [see Fig. 16.6(b)], and at the points of avoided frequency crossing the modes interchange their identities, i.e., Q-factors and field patterns. If however the wavelengths differ significantly, the coupling between the modes is weak, and the intensity of the PM super-modes is mostly localized in one of the cavities. Similar anti-crossing coupling of super-modes can be observed if the sizes of two photonic atoms are only slightly detuned from each other [15, 77, 78]. At the point of avoided frequency crossing, the super-mode eigenvectors represent a symmetric and an anti-symmetric superposition of the eigenvectors of the uncoupled system [41]. In the simplest case of an asymmetric double-cavity PM this effect causes formation of bonding (the symmetric superposition) and anti-bonding (the anti-symmetric one) PM super-modes at the point where both cavities are identical. If the microcavities are severely size-mismatched, their WG-modes couple with the frequency crossing scenario [77], which results in significant spoiling of their Q-factors.

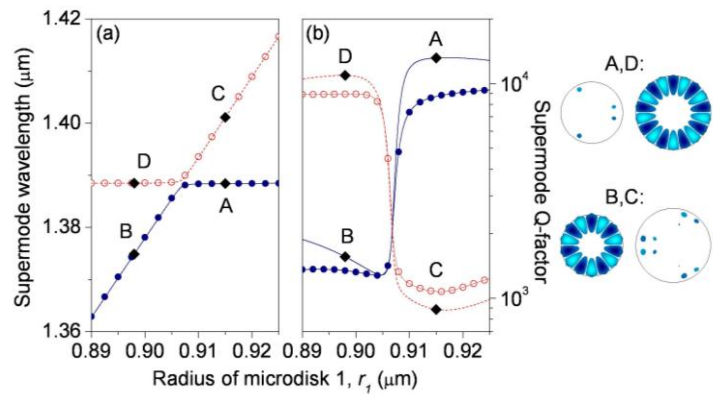

**Fig. 16.6** Wavelengths migration (a) and Q-factors change (b) of diatomic PM supermodes as a function of the radius of one of the disks ($r_2 = 1.1\mu m$, gap = 400nm, $n$=2.63). The insets demonstrate mode switching at the anti-crossing point (the modal near-field distributions shown in the upper (lower) insets correspond to the points labeled as A and D (B and C), respectively) [77].

However, weak coupling between severely detuned modes of photonic atoms can produce sharp asymmetric features in the PM transmission spectrum, similar to the Fano effect in atomic systems [79], which occurs as a result of the interference between a discrete autoionized state with a continuum [Fig. 16.7(a)]. Tunable Fano interference effect in an asymmetrical bisphere PM is illustrated in Fig. 16.7(c). Reshaping of the resonant peak of a single sphere by tuning it in and out of resonance with the second one can be observed. If the resonant frequencies of two spheres are tuned to the same value, CRIT phenomenon is observed resulting in appearance of a narrow transparent window [Fig. 16.7(c2)]. In the case of larger frequency detuning, this symmetrical transmission peak reshapes into a sharp asymmetric feature [Fig. 16.7(c3)]. This feature is a manifestation of a Fano-type resonance caused by coupling between a discrete energy state (a high-Q WG mode in one of the spheres) and a continuum of 'quasi' WG modes with non-circular shapes and severely reduced Q-factors [80, 81]. The field intensity distribution of the modes excited in bisphere PMs under such non-resonance condition is shown in Fig. 16.7(b).

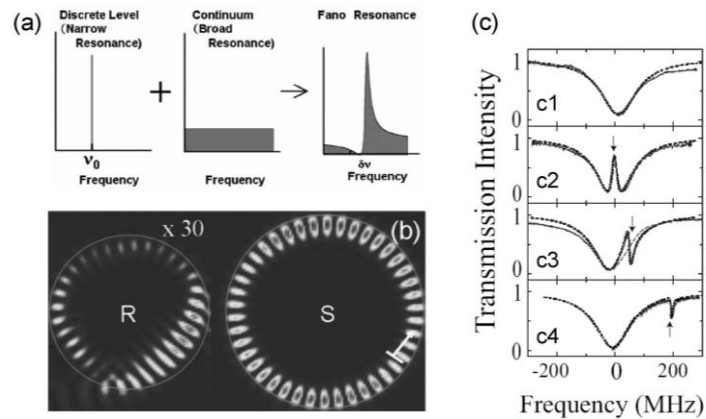

**Fig. 16.7** (a) Schematic illustration of the Fano effect in atomic transition: coupling between a discrete energy state and a continuum ionization level produces a sharply varying asymmetric structure [79]. (b) Intensity distribution in a bisphere PM in a weak coupling regime: coupling of a high-Q WG mode to a continuum of quasi-WG-mode states [81]. (c) Experimental (solid) and theoretical (dashed) transmission spectra of a single fiber-coupled microsphere (c1), and of a PM composed of two size-mismatched microspheres with the WG mode frequencies detuned from each other (c2-c4). The spectrum is centered at the WG-mode frequency of the first sphere, and the arrows indicate the resonance frequency of the second one [79].

Excitation of bonding and anti-bonding super-modes in PM structures also creates attractive and repulsive forces between photonic atoms [18, 82, 83]. The magnitude of this force scales linearly with the mode Q-factor, and also is a function of the coupling-induced frequency shift. The optical forces generated by different super-modes in a bisphere PM are plotted in Fig. 16.8(a) [82]. It can be seen that a bonding super-mode generates an attractive force, while an anti-bonding one generates a repulsive one. Fig. 16.8(a) also demonstrates that weak coupling between tightly-confined WG modes of higher radial number result in smaller forces. The forces are large enough to cause displacements of photonic atoms [82, 84] and may be harnessed for all-optical reconfiguration of photonic devices.

Coupling-induced optical forces give rise also to strong and localized optomechanical potential wells, which enable all-optical self-adapting behavior of PMs. This effect has been demonstrated in a diatomic PM structure composed of two vertically-coupled microring resonators, one static and one mobile (shown in Fig. 16.1(f)). The inter-resonator coupling can be modified by tuning the airgap $q$ between the microrings and causes WG mode splitting into bonding and anti-bonding PM super-modes [Fig. 16.8(b)]. If the PM structure is illuminated by the external laser light at a fixed frequency $\omega_L$, either a bonding or an anti-bonding mode is excited depending on the inter-cavity coupling strength. As a result, either attractive or repulsive forces are generated (Fig. 8(c)) leading to the creation of an effective potential between points (2) and (3) [18]. Resonant synthesis of potential wells can be used to obtain dynamical self-alignment between a PM super-mode and external laser light over a wide frequency range [18].

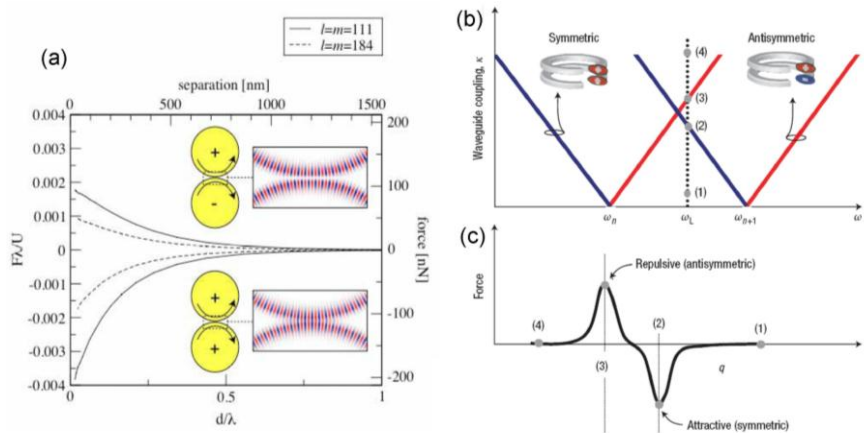

**Fig. 16.8** (a) Optical force generated by bonding and anti-bonding WG$_{lm1}$ super-modes in a bi-sphere PM as a function of inter-cavity separation [82]. (b) Bonding and anti-bonding mode splitting in a vertically-coupled double-ring PM as a function of the inter-disk coupling, and (c) forces versus ring-to-ring separation for a fixed wavelength excitation indicated by a dotted line in (b) [18].

## 16.3 Spectral engineering of photonic molecules

We have already observed that even in the simplest case of diatomic PM structures interference between PM super-modes can result in linewidth narrowing (Q-factor boost) and amplitude enhancement of certain super-modes (see Fig. 16.3, [50, 60, 77, 85]). More complex PM structures can be engineered with the aim to improve spectral characteristics of isolated optical microcavities and add new functionalities to microcavity-based optoelectronic devices. Many approximate techniques have been developed and applied to design and investigate PM structures, including perturbation methods [19, 82] and tight-binding approximation [10, 13]. These approaches, however, are only applicable to the analysis of non-degenerate modes, and should be used with caution in the simulations of coupled microresonators that support double- or multiple-degenerate WG modes, even in the case of weak coupling [54]. Accurate analytical methods such as multi-particle

Mie theory in 2D and 3D [50, 62, 86-88] and integral equation techniques [25, 89, 90] need to be used to properly reproduce the fine structure of the PM spectrum [54], account for intermixing between WG modes characterized by different values of the azimuthal quantum number [62, 77], and accurately consider coupling that occurs between non-neighboring cavities in the photonic molecule [8, 33-35, 72, 89]. It should be noted that extended multi-cavity coupling has been demonstrated both experimentally and theoretically not only in complex-shape PMs such as triangles, squares, etc. [8, 25, 33, 66, 89, 91], but also in slightly bent linear arrays of photonic atoms [66, 89]. The use of rigorous techniques becomes essential when the sizes of interacting optical atoms are on the scale of the wavelength. Mutual coupling between such atoms can significantly modify field distributions of WG modes, and causes significant detuning of super-mode frequencies from their single-cavity counterparts [25, 89].

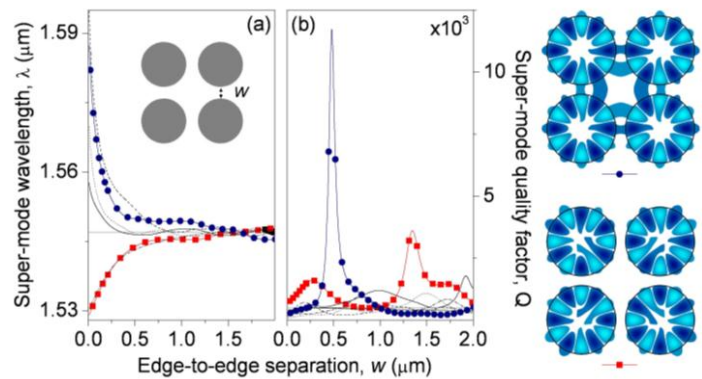

**Fig. 16.9** Shift and splitting of wavelengths (a) and change of Q-factors (b) of the TE-polarized $WGE_{6,1}$ modes in a square four-disk PM (disk radii 0.9μm, $n$=2.63) as a function of the inter-cavity gap width. The inset shows the magnetic field distribution of two PM super-modes whose Q-factors are boosted at certain values of the gap width (symmetry-enhanced super-modes) [33].

Among the advantages offered by PM structures over isolated microcavities is improved robustness against disorder. Rotational symmetry of microdisk or microsphere resonators can be broken by fabrication imperfections, resulting in splitting the double- or multiple WG-modes degeneracy and spoiling their high quality factors [92, 93]. Degenerate mode splitting in microcavities also occurs due to coupling to fibers and bus waveguides [94]. Optical response of PM structures can be more robust against disorder as the mode degeneracy is already lifted due to inter-cavity coupling [8, 77]. Furthermore, a quality factor of an optical mode is a very important parameter, which has to be maximized in order to enhance light-matter interactions in microcavities, to provide higher sensitivity of resonator-based optical sensors, to increase optical forces, etc. However, while the Q-factors of WG modes in microdisks and microspheres grow exponentially with the cavity size, the cavity free spectral range (FSR, spectral separation between neighboring high-Q modes) shrinks. Higher-azimuthal-order WG modes in large microcavities are also more sensitive to Q-factor degradation caused by the cavity surface roughness [93]. For many applications, however, such as low-threshold microlasers, single-photon sources, and optical biosensors, single-mode (or rather quasi-single mode) cavities with a narrow mode linewidth are required. Coupling cavities into PM structures offers a design strategy for the controlled manipulation of their optical spectra through destructive and constructive optical interference. Thus, only one mode (e.g. the one that has a resonant frequency within the material gain bandwidth) may be enhanced, and all the other modes (considered parasitic) are suppressed. An example of a wide-FSR quasi-single mode square-shape PM structure is shown in Fig. 16.9 [8, 25, 33]. Optical coupling between microdisks in such a configuration splits their degenerate WG modes into four non-degenerate super-modes with even/odd symmetry along the square diagonals and the x- and y- axes (termed as EE, EO, OE, and OO super-modes) as well as two double-degenerate modes. The modes with the highest Q-factors are the OE and OO super-modes (their near-field patterns are shown in the inset to Fig. 16.9). A dramatic 23-fold enhancement of the non-degenerate OE mode [Fig. 16.9(b)] in the optimal square-molecule configuration can clearly be observed. Note that all the other modes have significantly lower Q-factors at this point. Other symmetrical PM geometries such as equilateral triangles [33], circles and hexagons [8, 35, 95] can also be tuned to provide selective mode enhancements. Furthermore, the FSR of a PM structure can be increased by using the Vernier effect [36, 96, 97].

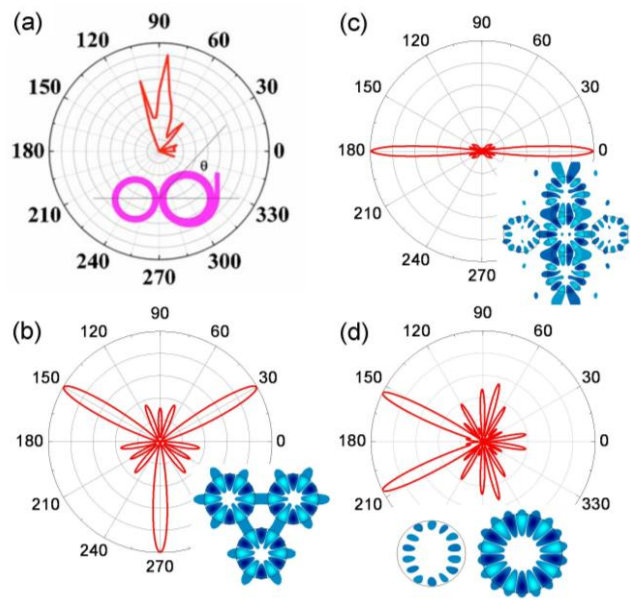

**Fig. 16.10** (a) Far-field emission pattern of the quasi-single-mode ring-spiral coupled PM structure [97]; far-field patterns and near-field distributions in coupled-microdisk PM structures shaped as (b) equilateral triangle [8], (c) asymmetrical cross [34], and (d) asymmetrical dumbbell [8, 77].

Many practical applications require directional emission from microcavity structures. One of the disadvantages of conventional WG-mode circular or spherical microcavities is their multi-beam emission patterns. However, proper engineering of inter-cavity optical coupling enables shaping far-field emission characteristics of PMs. One possible PM configuration featuring narrow-beam emission pattern is shown in Fig. 16.10(a) [97]. Here, emission directionality was achieved by coupling a microring resonator to a spiral-shape microcavity, which generates directional emission pattern owing to its highly asymmetrical shape [98]. Simultaneously, the Vernier effect was used to obtain a single-mode emission spectrum [97]. However, highly directional emission can also be achieved in PMs composed of circularly-symmetrical microdisk resonators [Fig. 16.10(b-d)]. Emission into four well-defined beams has been demonstrated in the simplest case of a double-disk PM [8, 36, 50, 57, 77]. The number of beams can be reduced (and directivity of the device improved) by using more complex PM configurations. For example, the EE-supermode supported by the triangular PM shown in Fig. 16.10(b) yields directional emission pattern with three narrow beams [8]. Emission directionality can also be improved by introducing position and/or size disorder in the PM structure. Emission patterns featuring just two narrow beams have been engineered in asymmetrical cross-shaped photonic molecules (Fig. 16.10(c) [34]) and asymmetrical diatomic molecules with cavity size mismatch [Fig. 16.10(d)] [8, 77]. Furthermore, emission into a pre-defined number of narrow beams can be achieved through the mechanism of multiple-nanojet formation in PM structures [99].

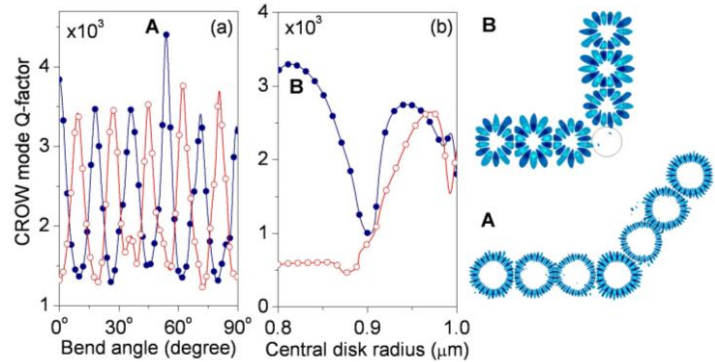

**Fig. 16.11** Quality factors change of two finite-CROW section super-modes forming a close doublet in the CROW spectra as a function of (a) the bend angle (disks radii 3.65μm, ε=2.5) and (b) the central resonator radius for the bend angle of 90° (disks radii 0.9μm, ε=7.0). The insets show the near-field portraits of high-Q super-modes in the optimally configured CROWs [89].

Optical microcavities arranged in linear chains can be used for light guiding by virtue of light propagating via evanescent coupling of microcavity modes. Such structures, termed coupled resonator optical waveguides or CROWs [10], provide new ways of controlling the group velocity of light on the optical chip and have been extensively studied in the literature (see e.g. [100] and references therein).

Here, we will focus on a specific application of CROWs, namely, a possibility of achieving very efficient or even complete transmission through sharp bends. In the pioneering paper introducing the CROW concept [10] it was suggested that if a CROW is composed of resonators that operate in an optical mode with an $n$-fold rotational symmetry, perfect transmission can be achieved through $2\pi/n$ bends. Efficient transmission through a microsphere CROW section with a bend angle of 90° has later been observed both theoretically [101] and experimentally [102]. Accurate integral-equation full-wave analysis of mode coupling and light transfer through CROW bends has revealed that this prediction was indeed valid for weakly-coupled microcavities supporting high-azimuthal-order WG modes [89]. This effect is illustrated in Fig. 16.11(a), which shows evolution of Q-factors of two nearly-degenerate high-Q $WG_{20,1}$ super-modes in the bent finite-size section of microdisk CROW with the change of the bend angle [89]. It can be seen that the Q-factors oscillate with the period $2\pi/n=18$ degrees. Furthermore, because the Q-values of two nearly-degenerate super-modes oscillate in anti-phase, this CROW can efficiently transfer light through $\pi/n$ bends. However, bending of CROW sections composed of smaller-size strongly-coupled microcavities can result in strong extended multi-cavity coupling [66] that may significantly disturb WG-mode field patterns and render the above estimation of the optimal angles inapplicable [89]. In this case, efficient transmission through any pre-defined angle can still be achieved by tuning the CROW structural parameters, e.g., the radius of the microdisk positioned at the bend [Fig. 16.11(b) [89]]. Furthermore, tuning sizes of individual resonators in linear CROWs may provide an additional degree of freedom for designing CROW band structures [103].

With the above examples we have demonstrated a variety of novel optical functionalities offered by PMs, both composed of identical microcavities and of microcavities with either slight or significant size mismatch. These functionalities together with higher design flexibility and tunability of PM structures over isolated photonic atoms pave the way for their use in various technological areas ranging from biotechnology to optoelectronics to quantum computing. In the next section, we will review some of the devices and applications based on spectrally-designed photonic molecules.

## 16.4 Devices and applications

Among the most promising potential applications of PM structures in integrated optics, signal processing and quantum cryptography is engineering of single-mode high-power microlasers and single photon sources. A single-mode microlaser can be engineered by optimally coupling two size-mismatched photonic atoms to yield selective enhancement of a single optical mode [36, 77, 97]. An example of such a device composed of two side-coupled microring resonators fabricated by coating glass fibers with thin layers of hybrid organic-inorganic materials is shown in Fig. 16.12(a). Proper adjustment of the microring radii selectively suppresses all the non-spectrally-overlapping modes of individual cavities and effectively increases the FSR of the resulting PM structure by means of the Vernier effect [36]. Furthermore, the structure asymmetry helps to improve the emission directivity, and thus boosts the microlaser collection efficiency (a portion of the emitted light that is redirected to the desired point or direction).

Other examples of PM microlasers include chain-like coupled-cavity structures [Fig. 16.12(b)] [67] and 2D arrays of coupled PC defect nano-cavities [Fig. 16.12(c)], [70]. In the structure shown in Fig. 16.12(b), coupling of the individual defect modes creates mini-bands within the bandgap of the InGaAsP/InP PC membrane, and stable single-mode lasing occurs on the mini-band mode with the lowest group velocity [67]. The 2D array of PC nano-cavities showed in Fig. 16.12(c) operates on a single high-Q super-mode and enables two orders of magnitude increase in the output power over a single-defect device, with a comparable low threshold power [70].

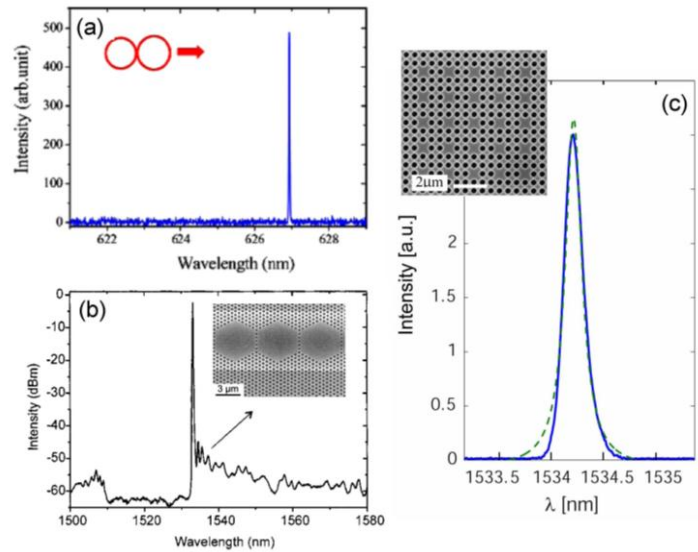

**Fig. 16.12** (a) Single-mode lasing from a diatomic PM structure composed of size-mismatched laterally coupled microring resonators ($n$=1.52, ring thickness 2μm, radii 115 and 125μm) [36]. (b) Laser spectrum of the PC coupled-cavity chain featuring quasi-single-mode lasing with side mode suppression greater than 42dB [67]. (c) Single-mode emission from a 2D array of coupled identical defect nanocavities in an InGaAsP PC membrane [70].

On the other hand, tunable dual- (or multiple-) wavelength laser sources are desirable in several applications such as two-wavelength interferometry for distance measurements [104], terahertz signal generation [105], and non-linear optical frequency mixing [106]. Two-wavelength laser emission has been successfully demonstrated in various types of vertical cavity surface emitting lasers (VCSELs) composed of two coupled microcavities containing multiple quantum wells [Fig.

16.13(a)], [107]. It has also been shown that to achieve stable dual-frequency lasing in such double-cavity sandwiches it is enough to pump only one of the cavities, whose emission then acts as an optical pump for the quantum wells in the other cavity [108, 109]. Coupling-induced splitting of the cavity optical modes in multi-atom photonic molecules leads to the appearance of multiple peaks in their lasing spectra, as demonstrated in Fig. 16.13(b) for the case of a microdisk PM structure.

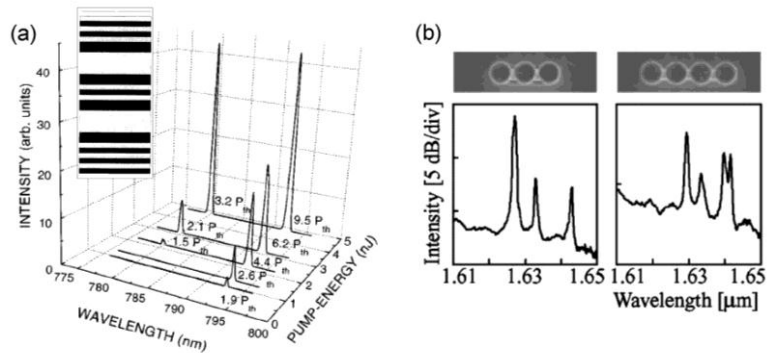

**Fig. 16.13** (a) Pump-energy dependence of time-averaged spectra of a VCSEL composed of two Fabry-Perot-type $Al_{0.2}Ga_{0.8}As/Al_{0.5}Ga_{0.5}As$ microcavities coupled via a common semi-transparent Bragg mirror shown in the inset [107]. (b) Top view and multi-mode lasing spectra of PM structures composed of several side-coupled 3μm-diameter GaInAsP microdisk resonators [53].

A possibility of manipulating the spectral response of PM structures by tuning the inter-cavity coupling strength also facilitated their use as multi-functional components for all-optical on-chip interconnection networks. By adapting the microwave circuit design principles, higher-order bandpass and add/drop filters can be engineered with cascaded microdisk resonators [Fig. 16.14(a)]. As shown in Fig. 16.14(b), higher-order optical filters offer flatter-top pass-bands with sharper roll-off than their single-disk counterparts, together with greater out-of-band rejection [26] (see also [5, 28, 30], Ch. 1 of [2] and references therein). On the other hand, filters with ultra-narrow passband (much narrower than the linewidth of each or the microcavities) can be engineered by making use of the EIT effect [27, 75] [see Fig. 16.5(b)].

Furthermore, the shapes of the transmission characteristics of the multi-cavity structures are very sensitive to the detuning of any of the cavities, making them attractive candidates for designing optical switches, routers, and tunable delay lines. For example, high-bandwidth optical data streams can be dynamically routed on the optical chip by tuning one or more microcavities in the cascaded high-order filter configuration out of resonance [110]. Detuning of the resonator frequency can be achieved all-optically by a pump laser [110] or electrically using a PIN diode. Alternatively, microcavities made of electro-optic crystals may be used [111]. The large thermo-optic effect of silicon and polymers makes possible tun-ing of resonant wavelengths of the microdisks by changing their temperature [112]. Selective opto-fluidic tuning of an individual cavity in a PM structure has also been demonstrated [113]. PM-based switches can be designed to feature a tailorable asymmetric Fano resonance response [see Fig. 16.7(c)] which yields high extinction ratio, large modulation depth and low switching threshold [114]. The linewidth of the narrow-passband filter based on the CRIT effect can also be manipulated by detuning the frequencies of the coupled photonic atoms away from each other [75]. It should finally be noted that all the above-discussed functionalities enabled by PMs can be realized not only with identical or size-mismatched cavities of the same type (e.g., coupled microrings) but also with completely different cavities (such as a microdisk coupled to a Fabry-Perot-type resonator formed by two partial reflectors in a bus waveguide) [75, 115].

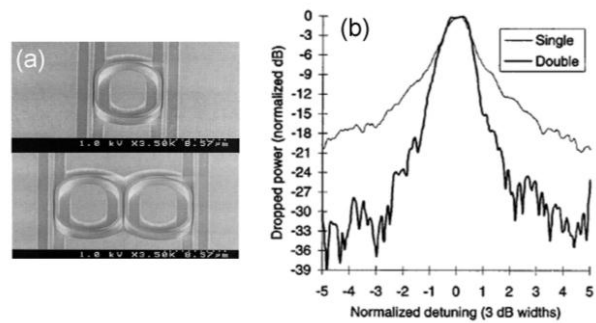

**Fig. 16.14** (a) Scanned electron microscopy (SEM) micrographs of first- and second-order Al-GaAs–GaAs microring racetrack resonator filters; (b) Measured transmission spectra at drop port of single- and double-resonator filters, with frequency scale normalized to 3dB widths [26].

Sharp and asymmetric Fano resonance features in the spectral characteristics of PMs can be useful for refractometric sensing applications, because the steeper slope between zero and unity transmission translates into a higher sensitivity to environment-induced resonant frequency shifts [31, 113, 115-117]. PMs can also exhibit large sensitivity to the change in the intrinsic loss in one of the optical cavities [60]. Another advantage offered by the PM structures is that different PM configurations exhibit distinct optical mode spectra, which may be considered PM optical fingerprints. These fingerprints can be used for tracking the PM response to the change of the ambient refractive index or to the adsorption of the molecules on the PM surface, even without the knowledge of the exact PM positions [24]. Such sensing scheme is highly promising for the in-situ multiplexed detection in the array format. Finally, using PMs supporting high-Q symmetry-enhanced super-modes (such as those shown in Fig. 16.9) can further improve sensitivity of microcavity-based optical biosensors [25]. Being collective multi-cavity resonances, super-modes of optimally-tuned PMs provide better overlap of the modal field with the analyte (see insets to Fig. 16.9) accompanied by the dramatic increase of their quality factors. Fig. 16.15 demonstrates the performance improvement offered by the PM-based sensor for the detection of the changes in the ambient re-fractive index over sensors based on individual microcavities with comparable Q-factors. PM structures with 4 and 16 disks can be tuned to resonate on symmetry-enhanced $WG_{4,1}$ super-modes, and feature larger frequency shifts caused by the change in the environment than the higher-radial-order mode of comparable Q-factor in an isolated larger-size microdisk.

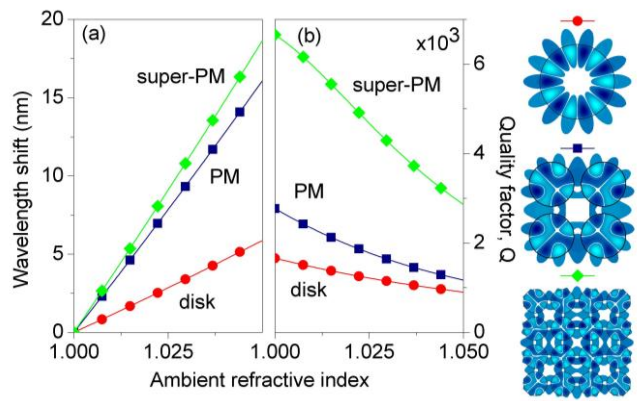

**Fig. 16.15** Comparison of the wavelength shifts (a) and Q-factors change (b) as a function of the ambient refractive index for a single 2μm diameter microdisk ($n = 2.63$) operating on a $WG_{7,1}$ mode, a square 4-disk PM, and a 16-disk super-PM structure operating on the symmetry-enhanced $WG_{4,1}$ super-mode ($n = 2.63$, radius = 0.67μm). Magnetic field portraits of the operating modes are shown in the inset [25].

The capability of coupled-microcavity structures to slow light also facilitates their use as ultra-compact optical gyroscopes for numerous practical applications that require detection and/or high-precision measurement of rotation. The principle of operation of optical gyroscopes is based on the Sagnac effect, i.e., the accumulation of the additional phase by an electromagnetic wave propagating in a moving medium [21, 22]. One possible configuration of a PM-based rotation sensor is shown in Fig. 16.16(a), and consists of several laterally coupled microdisk resonators. The relative phase difference between clockwise and counter-clockwise propagating signals in such a structure is a function of its angular velocity, and can be measured to detect the device rotation [22]. It can be seen in Fig. 16.16(a) that the sensitivity of the device can be enhanced by increasing the number of coupled microcavities.

It has already been mentioned that the sensitivity of the PM super-mode frequencies to the width of the inter-cavity coupling gap enables the use of PMs as optical parametric transducers for measuring small displacements [19]. Ultra-small sizes of PM structures offer unique opportunities for such ultra-sensitive applications as ponderomotive quantum nondemolition photon number measurements. If a position displacement in a PM structure is caused by the applied mechanical stress, after a proper calibration, it can be used as a stress sensor. A highly sensitive stress sensor design based on a PM structure composed of two photonic-crystal cavities vertically coupled via an airgap is shown in Fig. 16.16(b) [20].

The resonant frequencies of the bonding and anti-bonding super-modes supported by such a double-layer structure shift with the change in the airgap width [Fig. 16.16(b)]. Measurements of this wavelength shift can be translated into measurements of the applied stress. Furthermore, by geometrical tuning of the double-layer PM structure, selective enhancement of one of the super-modes may be achieved, leading to the improved sensor spectral resolution and sensitivity [20].

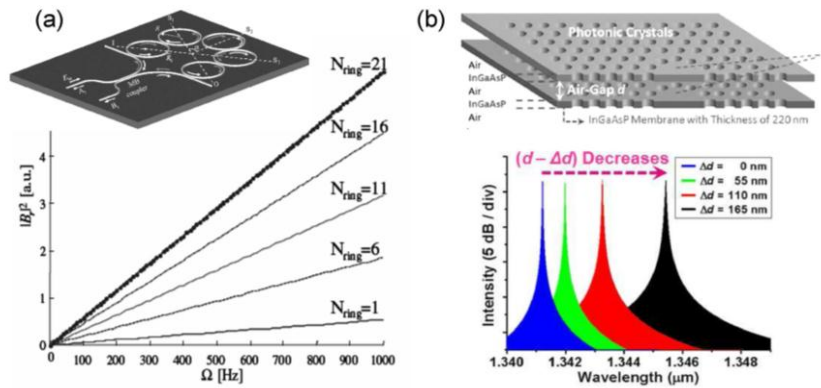

**Fig. 16.16** (a) Coupled-microring slow-light rotation sensor: the device schematic and the output signal intensity as a function of the PM angular velocity for structures composed of various numbers of ring resonators [22]. (b) Optical stress sensor: wavelength shift of the bonding supermode in a double-layered PC PM caused by the change of the width $d$ of the vertical airgap between coupled cavities [20]. The inset shows the schematic of the double-layer PM structure.

Spatial re-distribution of the optical modes inside PM structures occurring in the vicinity of the points of strong mode coupling with the avoided crossing scenario (see Fig. 16.6) offers exciting new prospects for switching optical intensity between coupled photonic atoms and realization of optical flip-flops. If the frequencies of two modes are detuned from each other (e.g. by cavity size mismatch), the optical fields of the resulting super-modes are mostly localized in different parts of the PM structure [15, 77, 78, 109]. Tuning the frequencies of individual cavities (e.g., by optical pumping) can result in the lasing bistability and mode switching in PMs [15, 78]. Bistable lasing achieved in a diatomic PM structure composed of two side-coupled GaInAsP microdisks [Fig. 16.1(c)] is demonstrated in Fig. 16.17. The PM lasing spectrum features two narrow peaks corresponding to the unresolved bonding ($S_1$) and anti-bonding ($S_2$) super-mode doublets [Fig. 16.17(a)]. The lasing characteristics of the highest-Q anti-bonding super-mode exhibits bi-stable behavior when the structure is pumped non-uniformly by an external laser [Fig. 16.17(b)] [15, 78]. Pumping changes the degree of microdisk frequency detuning and can be used to switch the optical intensity between microdisks.

Optical intensity switching can also be realized in more complex PM configurations consisting of several microcavities by selectively tuning the refractive index of a single cavity. For example, transmission through a CROW bend can be enhanced [see Fig. 16.11(b)] by tuning the resonator material parameters rather than the geometrical parameters. Another useful example of a switchable coupled-cavity structure is illustrated in Fig. 16.18 showing a branched CROW section that enables splitting of waves or pulses into two output waveguide ports or routing all the energy into a single port. As demonstrated in Fig. 16.18, switching between two high-Q supermodes of a branched CROW structure can be achieved by tuning the refractive index of the central disk [72].

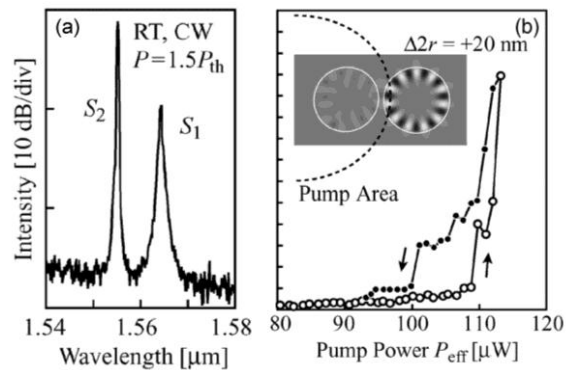

**Fig. 16.17** (a) Lasing spectrum of a size-mismatched twin microdisk PM shown in Fig. 16.1(c) [15, 78]. (b) Dynamics of the lasing characteristics of the anti-bonding super-mode ($S_2$) under pumping of the entire area of the smaller disk and only a small part of the larger one. Open and closed circles correspond to the measurements with increasing and decreasing pump intensities, respectively. The inset shows the super-mode field profile and the pumping area.

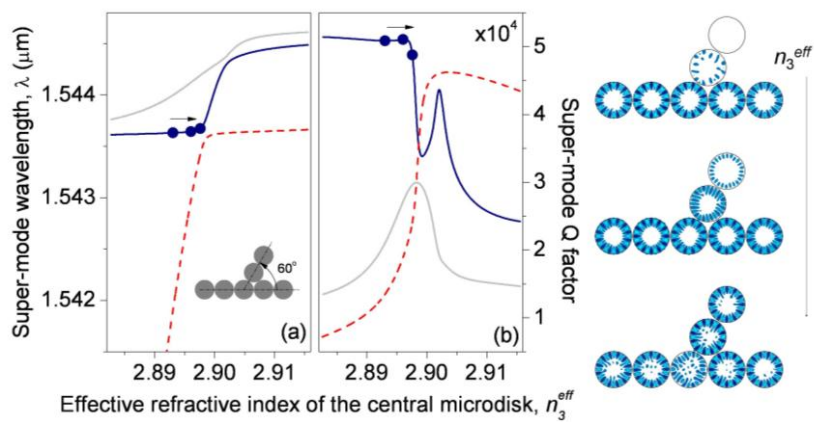

**Fig. 16.18** Switching in the branched CROW section. Change in the high-Q CROW super-modes resonant wavelengths (a) and Q-factors (b) with the change in the effective refractive index of the microdisk positioned at the CROW branching point. The inset shows a sketch of the branched CROW section ($n_{eff}$ = 2.9, diameters 2.8μm, airgaps 100nm). The magnetic field distributions in the CROW correspond to the three different values of the central disk refractive index marked with circles in (a) and (b) [72].

It should finally be noted that dynamical intensity switching between different parts of PMs can be used to coherently transfer excitation between quantum dots (QD) [118, 119] or quantum wells (QW) [14, 120] embedded in different cavities. Controllable interaction between bonding and anti-bonding PM supermodes and degenerate QW exciton states confined in separate cavities makes possible coupling between excitons over very large macroscopic distances [14]. Overall, the possibility to selectively address individual cavities in the PM structures doped with atoms or containing quantum wells/dots (e.g., by an external laser light) makes them very attractive platforms for simulating complex behavior of strongly-correlated solid-state systems [42, 43, 46, 121]. Controllable interaction of PM modes with atoms or QDs also paves the way to engineering devices for distributed quantum optical information processing applications [42, 122].

For example, PMs can form a useful platform for efficient optical probes that reveal an interplay between macroscopic coherence and interactions in strongly correlated photonic systems [44]. A possible design of an optical analogue of the superconducting Josephson interferometer (quantum-optical Josephson interferometer) is shown in Fig. 16.19(a) [44]. It consists of a central cavity with nonlinearity (e.g. caused by the light-matter interaction with an embedded QD), which is coupled to two other cavities driven by external lasers. Interactions between QDs and optical modes can be controlled by manipulating coupling between different cavity modes [122]. The interplay between interactions and inter-cavity photon tunneling can be explored by observing the light emitted from the central cavity under the change of the inter-cavity coupling strengths. A cross-over from the coherent regime (with optical intensity delocalized over all three cavities) to the strongly correlated regime can be detected.

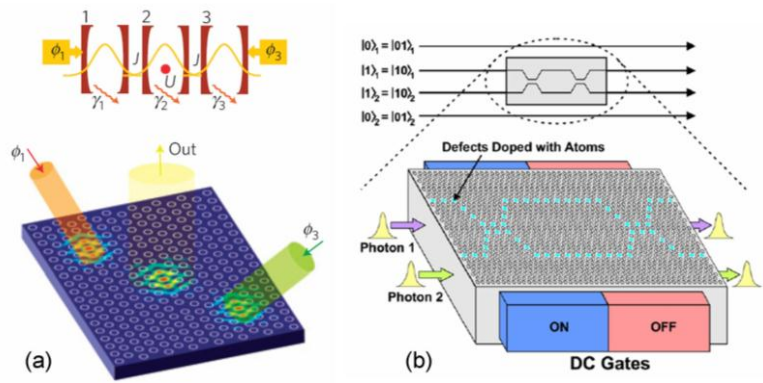

**Fig. 16.19** (a) Schematic of a quantum-optical Josephson interferometer with three coupled cavities and a possible device design with PC nanocavities, where the central cavity has an embedded QD or a QW strongly coupled with a high-Q optical mode [44]. (b) A schematic of a quantum logic gate composed of four CROWs representing two qubits, and a possible practical implementation of the device using coupled defects in a PC doped with atoms or QDs [45].

Furthermore, many other basic elements needed for quantum information processing, including state transfer [123, 124], entanglement generation [47, 125] and quantum gate operations [45, 126] can be realized in spectrally engineered PM structures. In particular, logic operations can be performed by exploiting strong coupling between atoms and optical super-modes in coupled-cavity PM structures. One possible design of a quantum gate is shown in Fig. 16.19(b). Two waveguides composed of coupled PC defect cavities doped with atoms or QDs are brought close to each other to enable optical coupling between them. These two waveguides represent two different qubits (units of quantum information). The dopant atoms or QDs can be tuned on- or off-resonance with propagating light by applying an external electric field. Interplay between (resonant) two-photon and (dispersive) one-photon transitions leads to phase shifts required both for single-qubit phase gates and two-qubit controlled-phase gates [45]. Mature modern nanofabrication technologies make possible practical realization of the above PM-based device designs and open new exciting opportunities in the fields of quantum cryptography and quantum information processing.

## Conclusions and outlook

Optical and light-matter interaction in PM structures composed of coupled microcavity resonators give rise to amazingly rich physics and can be harnessed for a variety of practical applications. PM structures already find their way into commercial applications for high bit-rate linear signal processing devices [2, 5, 12]. They also hold high promise for chip scale integrated sensing systems [116] and tunable optical buffers and delay lines [100, 110].

One of the emerging highly promising areas of photonic molecules research and application is cavity optomechanics. Coupling of optical and mechanical degrees of freedom in microcavities and photonic molecules provides unique mechanism of control and measurement of mechanical motion with micro- and nano-scale structures. Optical forces of switchable polarity that can be reversibly generated in coupled-cavity structures [18, 82-84, 129] offer exciting opportunities for the design of novel tunable lasers and filters, optomechanical signal processing, high-speed all-optical switching, wavelength conversion, and dynamic reconfiguration of optical circuits [130-132].

Furthermore, tunable interaction of light forces generated by photonic molecules with small biological objects paves the way to the development of ultrasensitive devices for mass, size and orientation spectroscopy. Recent experiments have demonstrated low-power optical trapping of nanoparticles in the orbit around a whispering-gallery-mode microsphere (Carousel trapping) [133]. Fluctuations in the microsphere resonance frequency were measured and used to estimate the size and mass of the trapped nanoparticle in the solution without binding it to the resonator surface. Coupled-cavity configurations are expected to provide additional degrees of freedom for detecting, capturing and manipulating biological nano-objects. For example, a PM sensor capable of measuring the dipole moment orientation of nanocrystals or molecules by tracking the dipole-induced spatial field distribution in the photonic molecule has already been theoretically demonstrated [134].

Finally, controllable manipulation of interactions between quantized PM optical modes and trapped atoms or embedded quantum dots opens up the possibility of using photonic molecules as quantum simulators of many-body physics and as building blocks of future optical quantum information processing networks [42].

## Acknowledgements

The author would like to thank Dr. Vladimir Ilchenko, Dr. Frank Vollmer, Dr. Sunil Sainis, Prof. Vasily Astratov, and Prof. Kerry Vahala for useful discussions. Support from the EU COST Action MP0702 "Towards functional sub-wavelength photonic structures" and from the NATO Collaborative Linkage Grant CBP.NUKR.CLG 982430 "Micro- and nano-cavity structures for imaging, biosensing and novel materials" is gratefully acknowledged.